\documentclass[10pt,psfig,a4paper,twocolumn]{article}
\usepackage{geometry}
\usepackage{graphics}
\usepackage{setspace}
\usepackage[scaled=0.9]{helvet}
\usepackage{color}
\usepackage[english]{babel}
\usepackage{psfrag}
\usepackage{epsfig}
\usepackage{graphics}

\begin{document}
\newcommand{\be}{\begin{equation}}
\newcommand{\ee}{\end{equation}}
\newcommand{\bestar}{\[}
\newcommand{\eestar}{\]}
\newcommand{\beastar}{\begin{eqnarray*}}
\newcommand{\eeastar}{\end{eqnarray*}}
\newcommand{\beq}{\begin{equation}}
\newcommand{\eeq}{\end{equation}}
\newcommand{\bea}{\begin{eqnarray}}
\newcommand{\eea}{\end{eqnarray}}
\newcommand{\dfrac}{\displaystyle\frac}
\newcommand{\disp}{\displaystyle}
\newcommand{\mbf}{\mathbf}
\newcommand{\dint}{\displaystyle\int}
\renewcommand{\u}{\underline}
\renewcommand{\o}{\overline}
\newcommand{\bi}{\begin{itemize}}
\newcommand{\ei}{\end{itemize}}
\newcommand{\bfig}{\begin{figure}[htb]\begin{center}}
\newcommand{\efig}{\end{center}\end{figure}}
\newcommand{\eqref}[1]{(\ref{#1})}
\newcommand{\eqm}[2]{~(\ref{#1}-\ref{#2})}
\newcommand{\eq}[1]{~(\ref{#1})}
\newcommand{\eqq}[2]{~(\ref{#1},\ref{#2})}
\newcommand{\eqqq}[3]{~(\ref{#1},\ref{#2},\ref{#3})}
\newcommand{\order}{{{\mathcal O}}}
\newcommand{\text}{\mbox}
\newcommand{\ie}{{\it i.e.}}
\newcommand{\eg}{{\it e.g.}}
\newcommand{\rd}{\color{red}}
\newcommand{\bl}{\color{blue}}
%define green to be xmgr's dark green (Green4), or not
\newcommand{\gr}{\color[rgb]{0.333333,0.752941,0.203922}}

\newcommand{\normal}{\mbox{\boldmath$\nu$}}
\title{Glow discharge in low pressure plasma PVD: mathematical model
  and numerical simulations}
\author{A. Speranza, A. Monti\\
  \multicolumn{1}{p{.7\textwidth}}{\centering\emph{Industrial
      Innovation Through Technological Transfer -- I$^2$T$^3$ Onlus}
    {\tt alessandro.speranza@i2t3.unifi.it}}\\
  L. Barletti, I. Borsi, L. Meacci\\
  \multicolumn{1}{p{.7\textwidth}}{\centering\emph{Dip. di Matematica
      ``U. Dini'' Universit\`a degli Studi di Firenze}}\\
  S. Fanfani\\
  \multicolumn{1}{p{.7\textwidth}}{\centering\emph{Galileo Vacuum
      Systems srl, Prato}}\\
}

%\author{A. Speranza\and L. Barletti\and L. Meacci\and
%  S. Fanfani\and I. Borsi\and A. Monti}
  
%\institute{A. Speranza\and A.Monti \at Innovazione. Ind. T. Trasf.
%  Tecnologico - I$^2$T$^3$ Onlus, Firenze\\
%  \email{alessandro.speranza@i2t3.unifi.it}
%  \and
%  L. Barletti\and L. Meacci\and I. Borsi\at 
%  Dip. di Matematica, Universit\`a degli Studi di Firenze
%  \and
%  S. Fanfani\at Galileo Vacuum Systems s.p.a., Prato}
\maketitle
\begin{abstract}
  In  this paper  we  analyze the  problem  of glow  discharge in  low
  pressure plasma in industrial plant,  for chambers of different shapes and
  various  working parameters, like  pressure and  electric potential.
  The model described  is based upon a static  approximation of the AC
  configuration   with   two   electrodes   and  a   drift   diffusion
  approximation  for  the  current density  of  positive  ions  and
  electrons. A detailed discussion  of the boundary conditions imposed
  is given, as well as the full description of the mathematical model.  

  Numerical simulations were performed for a simple 1D model and two
  different 2D models, corresponding to two different settings of the
  industrial plant. The simpler case consists of a radially symmetric
  chamber, with one central 
  electrode (cathode), based upon a DC generator. In this case, the steel chamber acts
  as the anode. The second model concerns a two dimensional horizontal
  cut of the most 
  common plant configuration, with two electrodes connected to an AC
  generator. The case is treated in a 
  ``quasi-static'' approximation. The three models show some common
  behaviours, particularly including the main expected features, such
  as dark spaces, glow regions and a wide ``plasma
  region''. Furthermore, the three shown models show some similarities
  with previously published results concerning 1D and simplified 2D
  models, as well as with some preliminary results of the full 3D case.     
  % In  spite of  the complexity of  the model  and the
  %strong non linearities introduced  in the coupling between the model
  %and  boundary conditions,  numerical  simulations in  1  and 2D  are
  %performed using  a FEM commercial algorithm.  The various algorithms
  %used  show good  stability and  the numerical  results  obtained are
  %comparable with previous results found  in the literature for the 2D
  %case. Furthermore, details of the distribution of ions and electrons
  %reveal  interesting  experimental  observations  and  so  far  shown
  %numerically only in 2D or cylindrically symmetric cases.
\end{abstract}

{\bf keywords}: PVD, glow discharge, surface coating, plasma, drift-diffusion
\section{The industrial problem}\label{sec:industrial}
Plasma PVD (Physical Vapour Deposition) is a widely used process,
applied for, particularly, in the coating industry. The process
basically aims at forming a thin coating layer on the surface of
different substrates. PVD is normally used as an alternative
physical process to classical wet chemical process, such as galvanic
baths.

\begin{figure*}[htb]
\begin{center}
\begin{picture}(0,0)%
\includegraphics{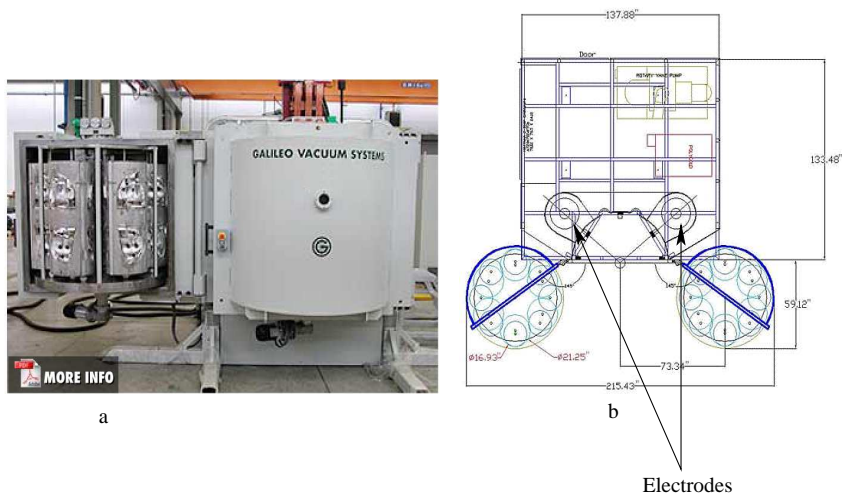}%
\end{picture}%
\setlength{\unitlength}{3522sp}%
\begingroup\makeatletter\ifx\SetFigFont\undefined%
\gdef\SetFigFont#1#2#3#4#5{%
  \reset@font\fontsize{#1}{#2pt}%
  \fontfamily{#3}\fontseries{#4}\fontshape{#5}%
  \selectfont}%
\fi\endgroup%
\begin{picture}(5885,3476)(496,-3762)
\end{picture}%
\label{Fig:camera_gvs}
\caption{a) Photograph of the industrial plant by Galileo Vacuum
  Systems s.p.a., objecto of the work. The headlights on the left hand
  side, have been already exposed to metalization, while the
  semi-chamber on the right hand side is closed and the process is
  active. b) Schematic representation of the PVD chamber (not the one
  produced by Galileo Vacuum Systems) as seen from
  the top. The chamber has two semi-chambers which can contain the
  headlights to treat, deposed on circular plates. The rectangular
  part above the two semi-chambers contains the electrodes and the
  vacuum pump.}
\end{center}
\end{figure*}

The case we particularly refer to concerns the
metalization of plastic headlights for automotive (Fig.~\ref{Fig:camera_gvs}), with a thin
aluminium layer, and its coating with thin polymeric film, such as
HMDSO (Hexa Methyldisiloxane). The HMDSO is used to protect the metalization from
external agents and ageing.

As it can be seen from Fig.~\ref{Fig:camera_gvs}-b, the industrial
plant consists of three main parts. The two semi-chambers visible at
the bottom of the figure, contain the headlights to be processed. 
The
two moving parts close alternatively on the fixed part of the plant
(rectangular in Fig.~\ref{Fig:camera_gvs}-b) as shown in the
photograph (Fig.~\ref{Fig:camera_gvs}-a). In this configuration, the
two moving semi-chambers contain circular plates disposed all
around the axis of symmetry, moving in planetary motion. The headlights
are deposed on the plates during the process. The fixed part of the plant, visible
as rectangle at the top of Fig.~\ref{Fig:camera_gvs}-b, contain the
electrodes (see below) and the vacuum pump with all the pipelining.
The industrial plant is normally made in two different
configurations. The configuration represented in figure, has two
cylindrical vertical electrodes and works on alternate current (AC), \ie, an alternate electric field is
applied to the electrodes in order to produce the glow discharge. In
the second configuration, one cylindrical electrode (normally the cathode) is
located in the center of the chamber and a direct electric potential (DC)
is applied to it. In this second configuration, the metallic chamber
itself works as the anode. 

The industrial process proceeds through three main steps
\begin{itemize}
\item Vacuum: one of the two mobile semi-chambers is locked against the
  fixed part, and vacuum ($\sim 10^{-3}$ mbar) is made
\item Metalization: aluminium filaments located all along the height
  of the mobile semi-chamber are heated via Joule effect. In such a
  low pressure atmosphere, aluminium vaporizes and a thin layer of
  metallic aluminium deposits onto the headlights surface. The more
  uniform and denser is the cloud of vaporized aluminium, the more
  uniform and thicker will be the metallic layer on the surface.
\item Polymer coating: a strong electric potential difference ($\sim$
  2-5 kV) is applied to the electrodes. In these conditions, the
  residual atmosphere transforms into plasma and a glow discharge
  appears. At the same time, a monomer flows through nozzles within
  the chamber, increasing the pressure up to 1-5$\times 10^{-1}$
  mbar. The plasma atmosphere induces the polymerization on the
  substrate of the headlights, and a thin polymer film covers the
  aluminium layer.   
\end{itemize}
In this work, we will neglect completely the first two steps of the
process, and will concentrate on the third one. In particular, we
will concentrate on the part in
which, in low pressure atmosphere, the electric potential is applied
to the electrodes (or to the cathode, in the configuration with one
central electrode) and the glow discharge is
induced~\cite{boeuf1988tdm,fiala1994tdh,HagHooKro00,reviewer:812,muta2002nil,roy2003mlp},
while the monomer flows and polymerization occurs. The aim of our
work is to model the glow discharge process, in order to gain information
concerning, \eg, the distribution of positive and negative charges, or
the electric field within the chamber, during the glow discharge. It is
clear, in fact, that HMDSO polymerizes under the influence of the
plasma atmosphere. However, it is not clear in what particular
conditions the polymerization is actually favored. In other words, our final
aim is to describe the plasma atmosphere all across the chamber, in order to
correlate its features, with the main features of the resulting
polymer film, such as thickness, homogeneity etc. With this
information, one could try to optimize the shape of the chamber and
the value of the main process
parameters, in order to have the best possible result, in the widest
part of the chamber. 
\section{The physical and mathematical model}
As mentioned in the previous section, this work focuses on the final
step of the industrial process, \ie, on the polymerization of the
HMDSO in low pressure plasma. In fact, our model concerns more the plasma
atmosphere in which the HMDSO polymerizes, rather than the process of
polymerization itself. The polymerization would be rather
difficult to model, as the process is strongly influenced by different
factors, not least the substrate on which the polymer film is formed
or its shape, besides the conditions of the plasma itself (see, for
instance, ~\cite{guo1998kdd,HegSchFis05}). For all these reasons, given that a realistic model of
polymerization would be rather difficult to achieve, and even more
complex to solve numerically (due to the strongly asymmetric geometry
of the domain), we focus on the analysis of the conditions in which
the monomer polymerizes while flowing across the chamber. Our
final aim
is to describe the conditions of the plasma, in terms of electric
field, density distribution of positive and negative charges etc., in
order to correlate these conditions with the main features of the
obtained polymer film.
\subsection{Hypotheses of the model}
Our mathematical model is based upon a set of simple assumptions,
aimed at simplifying the set of equations and, in parallel, the geometry of the domain. As far as
the chamber is concerned, we therefore assume 
\begin{itemize}
\item The steel chamber is perfectly cylindrical.
\item The electrodes (one or two, depending on the configuration of
  the plant) are perfectly cylindrical and not connected anyhow to
  the chamber. In other words, the electrodes ``float'' in the
  chamber, as they are generally shorter than the chamber itself.
\end{itemize}
In order to simplify the set up of the mathematical model, we then
make the following assumptions, concerning the plasma atmosphere:
\begin{itemize}
\item Thermal (and thermodynamic) equilibrium is assumed throughout
  the chamber. In this way, we assume a perfectly static atmosphere.
\item Plasma is made only of two species of charged particles:
  positive single ionized atoms (ions) and free electrons. Clearly
  this assumption is rather strong, especially in the final part of
  the process, \ie, when the monomer is flowing within the chamber and
  the polymer film is forming. When this happens, the atmosphere is made
  of a great number of species, such as monomers, ions,
  radicals, pieces of polymer chains and so on, however, this is the
  only possible starting point for any kind of modelling approach.
\item We always consider a stationary electric field, even in the case of
  an alternate potential difference applied to the electrodes. In this
  way we can neglect all the terms involving the variation of the
  electric field in the Maxwell equations (see below).
\end{itemize}
%The hypotheses above clearly reduce strongly the range of application
%of our model. The resulting model should therefore not be seen as a
%way to describe the actual industrial process as such. Our model
%should rather be seen as a tool to give som information to the
%experimentalists, and to give a first insight of the reasons why a
%certain feature of the resulting polymer film is particularly found
%only in some situations, and not in others. In other words, our model
%allows us to reduce the industrial problem to the combintation of
%three main parameters, such as the atmospheric pressure, the potential
%applied to the electrodes and the global geometry of the plant
%(especially for what concerns the shape of the electrodes and their
%position within the chamber), so that we can try to correlate them
%with the actual result of the real process.
%
Moreover, we will make a further assumption, concerning, as we will
see in later sections, the boundary conditions to impose to our set of
equations: we assume that all the steel parts composing the chamber
and the electrodes, are ``perfectly absorbing'', \ie, all the charged
particles hitting the surface of any steel part, go back into the
chamber as neutral particles. In other words, charged particles are
never reflected into the chamber, as they are.

Before proceeding to the description of the mathematical model, let us
go back for a moment to the assumption of a static electric field
mentioned earlier, in spite of the choice of the
chamber configuration with one (DC generator) or two (AC generator)
electrodes. In fact, this assumption is reasonable, as, even in the
case of two electrodes, the frequency of the applied field is of the
order of 100 kHz. Given that the radius of the chamber is normally of
the order of 1 m, and that the wavelength of the applied field is of
the order of 3 km, it turns out that, within the extension of the
chamber, the electric field can be considered uniform. In other words,
the chamber is sufficiently small to allow us the use of a
``quasi-static approximation'', \ie, to neglect all the
possible delay effects due to the variation of the electric
field within it. 
%%%%%%%%%%%REFEREEE 1 - P.3%%%%%%%%%%%%%%%%%%%%%%%%%%%%%%%%%%

As we will shortly see in the following sections, in spite of the
drastic simplifications of the physical problem that we made in order to
solve numerically the mathematical model, our results turn out to
correlate fairly well with observations and direct measures. In fact,
although the real atmosphere consists of many types of ions and
radicals, resulting from the polymerization process, observations
show, as in simpler atmosphere, the formation of a ``dark space''
surrounding the electrodes, confined by a bright region: ``the
glow''. As we will see in the following
sections, in our numerical results the two regions are clearly
evident. The two regions differ in particular in the values of the number density of charged particles
and electric field intensity. In other words, in spite of the high
degree of approximation of our model, our numerical results turn out to give some reasonable
insights of the actual problem, marking an important step forward in
the interpretation of observations and direct measures, aimed at the
optimization of the industrial setting.  
%%%%%%%%%%%%%%%%%END OF REFEREE PARAGRAPH%%%%%%%%%%%%%%%%%%%%%%
%In other words, the assumption of a static electric field
%corresponds to freezing the conditions of the chamber at a certain
%time $t$. The full dynamic model could be reconstructed by summing
%many steady models like ours, taken at different times, \ie, with a
%different potential difference applied to the electrodes. 
%
\subsection{Set up of the equations}
Once the physical model has been set as described in the previous
section, the set of equations to solve turns out to be rather
simple. As mentioned, we assume a steady atmosphere and thermal
equilibrium across the chamber. Furthermore, we assume a steady
electric field applied to the electrodes. In these hypotheses, the
electric potential must obey the non homogeneous Poisson equation,
\ie, 
\begin{equation}\label{eq:Poisson}
\Delta V=-\frac{e}{\epsilon}\left(n_i-n_e\right)
\end{equation}
where $\Delta$ is the Laplace operator, $V$ is the electric potential,
$e$ is the electron electric charge ($\simeq$ 1.6 $10^{-19}$ C), $\epsilon$ is
the absolute dielectric constant of the atmosphere and $n_{i},\ n_e$ are
the number densities of ions and electrons respectively. For the
equation above, one must set boundary conditions which, in this 
case, turn out to be straightforward Dirichlet boundary conditions, as
it will be shown in the following section.

\begin{figure*}[htb]
\begin{center}
\begin{picture}(0,0)%
\includegraphics{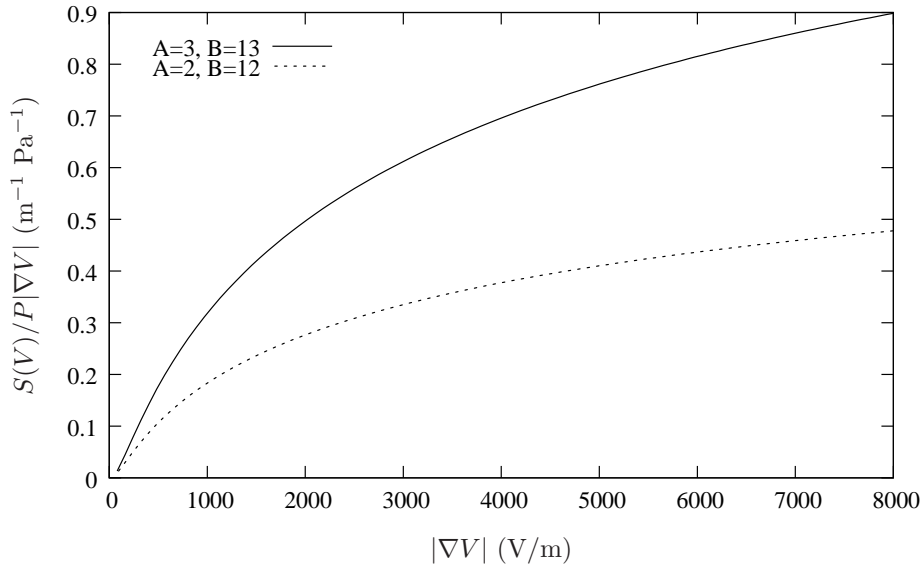}%
\end{picture}%
\setlength{\unitlength}{3947sp}%
\begingroup\makeatletter\ifx\SetFigFont\undefined%
\gdef\SetFigFont#1#2#3#4#5{%
  \reset@font\fontsize{#1}{#2pt}%
  \fontfamily{#3}\fontseries{#4}\fontshape{#5}%
  \selectfont}%
\fi\endgroup%
\begin{picture}(5726,3542)(1471,-3995)
\put(4550,-3935){\makebox(0,0)[b]{\smash{{\SetFigFont{10}{12.0}{\familydefault}{\mddefault}{\updefault}$|\nabla V|$ (V/m)}}}}
\put(1606,-1974){\rotatebox{90.0}{\makebox(0,0)[b]{\smash{{\SetFigFont{10}{12.0}{\familydefault}{\mddefault}{\updefault}$S(V)/P|\nabla V|$ (m$^{-1}$ Pa$^{-1}$)}}}}}
\end{picture}%
\caption{First Townsend coefficient rescaled by pressure and electron
  mobility as
  defined in Eq.~\eqref{eq:Townsend}, as a function of $|\nabla V|$. 
The two sets of coefficients $A=2-3$ cm$^{-1}$
Torr$^{-1}$ and $B=12-13$ (V/cm Torr)$^{0.4}$ correspond roughly to the
data published in~\cite{ward2004ccf} for He.}
\label{fig:SV}
\end{center}
\end{figure*}

As far as ions and electrons current densities $\mbf{J_{i,e}}$ are
concerned, in the hypothesis of a continuous model (see below), they
obviously must obey the
stationary continuity equation, \ie,
\begin{eqnarray}
\nabla\cdot\mbf{J_i}&=&\mu_e n_e S(V)\label{eq:ion-current}\\
\nabla\cdot\mbf{J_e}&=&\mu_e n_e S(V)\label{eq:electron-current}
\end{eqnarray}
where $\mu_{e}$ is the electrons mobility coefficient and
$S(V)$ is the ionization frequency, which is obtained from the
Townsend formula~\cite{BoePit95,townsend1915electricity,ward2004ccf}
\begin{equation}\label{eq:Townsend}
S(V)=A\exp\left\{-B\left(\frac{P}{\left |\nabla V\right |}\right)^{0.4}\right\}P\left |\nabla V\right |
\end{equation}
The term on the right hand side of Eqs.~\eqq{eq:ion-current}{eq:electron-current}, $\mu_en_eS(V)$,
basically expresses the number density of ion/electron pairs 
formed through a collision between a neutral particle and a free
electron, per unit time. As obvious, this term depends on the pressure $P$
and the electric field module $|\nabla V|$. Note that, compared to the
standard notation (\eg,~\cite{BoePit95,roy2003mlp}), we pulled the electron
mobility out of the definition of the Townsend coefficient, in order to
make the typical scaling of the equations more evident. The two coefficients $A$
and $B$ in the equation above, are obtained by a fit of
experimental data, and depend on the composition of the atmosphere
(see,
\eg,~\cite{posin1936tca,sanders1932vtc,sanders1933mtc,townsend1915electricity,ward2004ccf}). Typical
values of the two coefficients range between 4 and 60 cm$^{-1}$ Torr$^{-1}$,
for $A$, and between 14 and 36 (V/cm Torr)$^{0.4}$, for $B$. In the
following, we made a fairly conservative choice, in order to control
more easily the numerical calculations and used $A$ between 2 and 3
cm$^{-1}$ Torr$^{-1}$ and $B$ between 12 and 13 (V/cm Torr)$^{0.4}$,
comparable with the values for He reported by Ward~\cite{ward2004ccf}
and with the values used by Boeuf and Pitchoford~\cite{BoePit95}.
As shown in Fig.~\ref{fig:SV}, the first Townsend coefficient is
strongly dependent on the choice of the set of coefficients $A$ and
$B$. However, as the electric field grows, and $S(V)/P|\nabla V|$
settles on a constant value, the frequency of ionization $S(V)\mu_e$ becomes linear with $|\nabla V|$.
%With this choice for the set of parameters, the Townsend coefficient,
%rescaled by the electron mobility as defined above, for high values of
%the electric field grows almost
%linearly with it, as $S(V)/P|\nabla V|$ becomes constant, as shown in Fig.~\ref{fig:SV}.

The current density $\mbf{J_q}$ of charged particles $q$ ($i$ or $e$)
in Eqs.~\eqq{eq:ion-current}{eq:electron-current} is modeled through a simple
drift-diffusion equation, \ie,
\begin{equation}\label{eq:drift-diffusion}
\mbf{J_q}:=-\mbox{sgn}(q)\mu_qn_q\nabla V - D_q\nabla n_q
\end{equation} 
where $D_q$ is the diffusivity coefficient and is linked to the charge
carrier mobility via the so called Einstein's formula
\[
D_q=\dfrac{\kappa_{\text B} T_q}{e}\mu_q
\]
where $\kappa_{\text B}$ is the Boltzmann constant, and $T_q$ is the
temperature of the charged particle. In the following, we made the assumption that ions are in thermal equilibrium
with the neutral gas, therefore $T_i=T=300$ K, while the
electron temperature was set to $T_e=10000$ K throughout our domain,
corresponding to an average energy $E_e\sim 0.4$ eV. The electrons and
ions mobility are taken, coherently with the choice we made for the
first Townsend coefficient, as $\mu_e=$ 1333 m$^2$ V$^{-1}$s$^{-1}$
($\mu_eP$=10$^6$ cm$^2$V$^{-1}$s$^{-1}$Torr) and $\mu_i=$10.66 m$^2$ V$^{-1}$s$^{-1}$
($\mu_iP$=8$\times 10^3$ cm$^2$V$^{-1}$s$^{-1}$Torr) respectively, at a
pressure of 10 Pa, as reported by Ward~\cite{ward2004ccf}. 
  
Our set of equations therefore consists of
Eqs.~\eqref{eq:Poisson}-~\eqref{eq:ion-current}
and~\eqref{eq:electron-current}, with the use of
~\eqref{eq:drift-diffusion}, in the three unknowns $V$, $n_i$ and $n_e$.
\subsection{Comments on the model}
%
%%%%%%%%%%%%%%%%%%%%%REFEREE 2 - Major revision%%%%%%%%%%%%%%%%%%%%%
As shown, our model is based upon the heuristic assumption, to be
verified {\em a posteriori}, that a fluid model is suitable throughout
the whole domain. Clearly, this assumption will have to be checked
against the parameters used to define the conditions of our plasma.
It is, in fact, well known, that in a very low pressure
plasma a collision-free sheath could be formed in contact with an
absorbing or a neutral wall~\cite{Bohm49,TonLan29}, effectively
confining the neutral plasma. As the hypothesis of high collisionality
is not satisfied, a continuous model may not be applicable to this case.  
%This assumption, at first sight, seems to contradict
%the well known phenomenon of the formation of a collision-free sheath
%in the immediate vicinity of an absorbing wall, such an electrode or a
%neutral wall (see \eg, ~\cite{Bohm49,TonLan29}). 
Within the sheath region, electrons are almost absent and $n_i\gg
n_e$. Furthermore, the sheath is confined by a
transition region (the pre-sheath) where the plasma is almost exactly
neutral (\ie, $n_i\simeq n_e$). The sheath is stable when the so called {\em Bohm
  criterion} is satisfied, \ie, when the ions drift velocity is larger
than some critical speed, corresponding, as shown later~\cite{Allen76,AndSta70},
to a Mach surface for the
ions flux. As mentioned, when the conditions for the formation of the sheath are
satisfied, a purely continuous model cannot be used, as the assumption
of high collisionality is not satisfied within the sheath and the
pre-sheat. In this case, one possible approach is
to develop a fluid model that connects, in the limit, with the
sheath model (see, \eg,~\cite{FraOck70,WanWen99}). However, the
conditions for the formation of a stable sheath are not completely clear
yet. It was
shown that the Bohm criterion results, in fact, in a
sufficient but not necessary condition for the sheath
formation~\cite{Riemann97,Valentini96}. 
It was also shown by Valentini~\cite{Valentini96} and Riemann~\cite{Riemann97}, that such
a criterion has to be strongly modified when collisions are
present. In other words, a sheath could be formed even in collisional
plasma, although its characteristics are quite different in this
case. Godyak and Sternberg~\cite{GodSte02} showed that 
the model upon which the formation of such a sheath is described, is
intrinsically inconsistent, suggesting that a modification on the
basic assumptions should be made. Valentini~\cite{Valentini00}, on the
other hand, showed that in a discharge plasma, \ie, a
slightly ionized, low pressure plasma, similar to ours, the sheath can
still be formed, under appropriate conditions. However, the whole
sheath concept, in particular in front of the cathode, should be somehow
redefined. In fact, in these conditions, a fluid model can usually be used
throughout the whole domain, as collisions are frequent and important, 
even though, the electrode fall still region turns out to
show very peculiar characteristics, compared to the neutral plasma
region. In order to take into account of the peculiar behaviours of
that region, authors chose to use
different methods. Hagelaar et al., for
instance~\cite{HagHooKro00}, used boundary conditions to take into
account of the different behaviour of the fast electrons generated
within the cathode fall region. Surendra et
al.~\cite{SurGraJel90}, coupled a fluid model (similar to ours) with
another model for the fast electrons. On the other hand, other
authors~\cite{Boeuf87,boeuf1988tdm,fiala1994tdh,GogNicSaw89,Graves87,MeyKre90,RicThoSaw87,roy2003mlp,SchEmm88,ZawNajCon86}
used a fluid model throughout the whole domain,
investigating its range of applicability, and
choosing boundary conditions in order to have a tractable and yet physically
meaningful model. Boeuf and Pitchford~\cite{BoePit95}
argued that a fluid drift-diffusion model can normally be used with
values of the pressure of the order of 100 mTorr ($\sim$ 13 Pa). In
fact, the same approach was also used for RF glow discharge
models~\cite{BarCotElt87,Boeuf87,Graves87,GraJen86,PasGoe93,RicThoSaw87}, 
as long as the number of collisions remains high enough, \ie, as long as the
applied pressure is sufficiently high. The threshold value for the pressure normally grows from
100 to 500 mTorr ($\sim$ 67 Pa), for applications of the fluid model to an RF
discharge model. However, the argument, in both cases, the AC and DC, is based upon the fact
that, for pressures above some values, the ions mean free path
is normally the smallest length scale, compared to, for instance, the
electrons Debye length, the typical length scale of the model, and so
on. In these conditions, collisions can be considered frequent enough to allow for a continuous
model to be physically meaningful.

A similar argument applies exactly also to our case. As mentioned above, we
chose a pressure value of 10 Pa ($\sim$ 75 mTorr), which is very close
to the lowest value mentioned, for instance by Boeuf and Pitchford~\cite{BoePit95}. In these
conditions, the ions mean free path $\lambda_i$ turns out to be of the
order of 1 mm. This value should be compared with the typical
length scale of the chamber ($L=1$ m) and, for instance,
the electrons Debye length $\lambda_{\mbox{D}}$. In our case,
the electrons Debye length turns out to be
$\lambda_{\mbox{D}}=L/\sqrt{\tilde{T_e}/\tilde n_e}$,
where $\tilde{T_e}=T_e/T$ and $\tilde n_e$ is the non dimensional electrons number
density (see next section). Since we set $\tilde{T_e}\sim 30$ and from our numerical
results, we have at most $\tilde n_e\sim 10^6$, we get that
$\lambda_{\mbox{D}}> 10$ mm. In any case, near the electrodes, where
$\tilde n_e$ decays down to 10$^3$, we have $\lambda_{\mbox{D}}\sim 0.1$ m. This particular result
indicates~\cite{Riemann91} that the electric field may not be strong
enough to confine the ions in a very narrow region in contact with the
electrode. Instead, collisions allow the ions to diffuse further 
inside the bulk region. This general consideration is supported by our
numerical results. As we will see in the following sections, the cathode
sheath region, shows a dominance of ions, as it should be, but a still
large presence of electrons ($n_e>10^8$
m$^{-3}$). Furthermore, our  
sheath is confined by a region with a high net charge density, and
not by a neutral region such as the one assumed to confine the sheath
region in the collision-less plasma. In
fact, as already predicted by Valentini~\cite{Valentini00}, in our
case, the net
charge density reaches a maximum before decaying rapidly to zero in the neutral
plasma region, coherently with the assumption of a highly collisional
plasma. Sheridan and Goree~\cite{SheGor91} showed that the
non-collisional and the collisional regimes connect smoothly through
an intermeidate regime with the relavant index being, once again, the
collision parameter $\alpha=\lambda_{\mbox D}/\lambda_i$. The authors argued that the
collisional regime extends in the region $\alpha > 0.1$. For values of
$\alpha$ of the order 10, like in our case, the regime is already
highly collisional and the fluid model should be fully
acceptable. 

In spite of the low pressure, we therefore conclude that our plasma shows a still
strongly collisional regime that allows us, although close to the
limiting case, to use a continuous model throughout
the domain. However, even when a smooth connection between the the
numerical results of the fluid model and an asymptotic analysis of the
boundary layer was attempted~\cite{WanWen99}, the two solutions showed
an appreciable difference only at a very short length-scale which are
far beyond our aims. Clearly, the argument of high collisionality applies only since we chose
to focus on the final step of the industrial process (that we referred
to as ``Polymer coating'' in Sec.~\ref{sec:industrial}). The
pressure there grows rapidly between 10$^{-3}$ mbar to 0.1-0.5 mbar as the monomer flows through the chamber and
polymerization occurs. The same argument, on the other hand, would not apply if we wanted
to describe the conditions of the plasma atmosphere at the very start of
the process, \ie, when the pressure is still 100 times lower, and the
atmosphere consists only of the residual air present in the chamber.
%%%%%%%%%%%%%%%%%%% END OF REFEREE PARAGRAPH%%%%%%%%%%%%%%%%%%%%%%%%%%% 
%
\section{The problem of boundary conditions}
To the set of equations described in the previous section, one has to
impose boundary conditions on each one of the steel parts which
confine the problem domain, \ie, the two electrodes and the chamber
surface. As mentioned in the previous section, to the Poisson equation
for the electric potential Eq.~\eqref{eq:Poisson}, one can simply
impose Dirichlet boundary conditions. For instance, defining $V_C$, the
potential of the steel chamber, as the reference potential, and
redefining all the potentials as potential differences $\tilde
V=V-V_C$ and swapping $V$ and $\tilde V$, we can set 
\begin{eqnarray}
\label{eq:Poisson_BC}
  V=&V_0 &\  \mbox{on the anode}\nonumber\\
  V=&-V_0 &\  \mbox{on the cathode}\nonumber\\
  V=&0 &\  \mbox{on the steel chamber}
\end{eqnarray}
In this way, the potential difference between
the two electrodes is set to be 2$V_0$. The
potential difference with the chamber is, respectively, $V_0$ and
$-V_0$ for the anode and the cathode. 

%While the boundary conditions for the Poisson equation are rather
%straight forward, and follow simply from the choice we made for the
%physical model, 
There is no such a simple way to express boundary
conditions for the number densities $n_i$ and $n_e$, unless we
make some further assumptions on the actual interaction between
charged particles and the electrodes or the chamber. Let us therefore
analyze the condition to impose on the different parts of the
plant, for the two species of charged particles. For the ions following, \eg, the arguments by Hagelaar et al.~\cite{HagHooKro00}, one can
make the following assumptions:
\begin{itemize}
\item The cathode absorbs ions, in the sense that any ion hitting the
  cathode surface, captures a free electron emerging on the surface
  and is reflected into the chamber as a neutral particle. We assume,
  as mentioned earlier, that the cathode is a ``perfect absorber of
  ions'' in the sense that none of the ions hitting the surface is simply
  reflected into the chamber. In these conditions, the density flux
  $\mbf{J_i}$ reduces to just the drift term, while the diffusion
  vanishes, \ie, we set a simple Neumann boundary condition such as 
  \begin{equation}\label{eq:ion_cathode}
    \nabla n_i\cdot\normal=0
  \end{equation}
  where $\normal$ is the normal versor pointing inward the cathode surface
  (\ie, outwards from our domain)
\item The anode does not emit (or absorb) ions, thus the net inward flux
  across its surface must vanish. In other words, on the anode we set
  a mixed boundary condition of the type
\begin{equation}\label{eq:ion_anode}
  -\mbf{J_i}\cdot\normal:=-(-\mu_i n_i\nabla V-D_i\nabla n_i)\cdot\normal=0
\end{equation}
where $\normal$ is the normal versor pointing inward the anode surface
(\ie, outwards from our domain)
\item Following this argument, the steel chamber will have to behave
  in a way which is similar to the anode or the cathode, depending on
  the direction of the electric field (and thus, of the ions
  flux). When the electric field $-\nabla V$ ``enters'' the
  surface, \ie, has the same sign as the outwards normal versor, the
  steel chamber behaves like a cathode. On the other hand, when the
  electric field $-\nabla V$ is outward the surface, \ie, entering our
  domain, and therefore anti-parallel to the outward normal versor,
  the steel chamber behaves like the anode. In compact form, we can
  write this condition as
  \begin{eqnarray}\label{eq:ion_chamber}
    \mbf{J_i}\cdot\normal&:=&(-\mu_i n_i\nabla V-D_i\nabla
    n_i)\cdot\normal=\nonumber\\
&=&-a_i\mu_in_i\nabla V\cdot\normal
  \end{eqnarray}
  where $a_i=1$ (cathode-like) if $-\nabla V\cdot\normal>0$ and $a_i=0$
    (anode like) otherwise
\end{itemize}
A symmetric argument applies for the electrons. In this case, however,
the roles of anode and cathode are inverted, thus
\begin{itemize}
\item The anode absorbs electrons, in the sense that any electron,
  hitting the anode, is captured by it. Thus, the diffusion term of the
  electrons flux vanishes on its surface, \ie,
\begin{equation}\label{eq:electron_anode}
  \nabla n_e\cdot\normal=0
\end{equation} 
\item On the cathode surface, the electrons inward flux is
  due to thermal and secondary emission, thus
\begin{eqnarray}\label{eq:electron_cathode}
-\mbf{J_e}\cdot\normal&:=&-(\mu_e n_e\nabla V-D_e\nabla
n_e)\cdot\normal=\nonumber\\
&=&\gamma_i\mbf{J_i}\cdot\normal+\frac{1}{4}n_e v_{\mbox{th}}
\end{eqnarray}
where $\gamma_i$ is the secondary emission coefficient (0.1 in our case) and
$v_{\mbox{th}}=\sqrt{8\kappa_{\mbox{B}}T/\pi m}$ is the thermal velocity
and $m$ is the electron mass.
\item As in the case of the ions, the steel chamber behaves
  differently, depending on the sign of the electric field, \ie, 
\begin{eqnarray}\label{eq:electron_chamber}
  \mbf{J_e}\cdot\normal&:=&(\mu_e n_e\nabla V-D_e\nabla
  n_e)\cdot\normal=\nonumber\\
&=&a_e\mu_en_e\nabla V\cdot\normal
\end{eqnarray}
where $a_e=0$ (cathode-like) if $-\nabla V\cdot\normal>0$ and $a_e=1$
    (anode like) otherwise
\end{itemize}
\subsection{Final setting of the model}
In order to make our result as general as possible, let us
define new non dimensional variables $u=V/V_0$ for the potential,
$\tilde n_i=n_i/n_0$ and
$\tilde n_e=n_e/n_0$, where $V_0$ and $n_0$ are scaling factors to be
chosen appropriately. Let us also rescale all the coordinates by the
typical length scale of the chamber ($L=1$ m). Inserting the new variables into
Eqs.~\eqm{eq:Poisson}{eq:electron-current} and swapping $\tilde n_i$ and
$\tilde n_e$ with $n_i$ and $n_e$, respectively, we can set
$V_0=\kappa_{\mbox{B}}T/e\ (\sim 10^{-2}$V) and
$n_0=\epsilon\kappa_{\mbox{B}} T/e^2L^2\ (\sim 10^5 \text{m}^{-3})$. With this
notation, we get our final model in non dimensional form as
\begin{eqnarray}
-\Delta u= n_i - n_e\\
\nabla\cdot\left(-n_i\nabla u-\nabla n_i\right)=\frac{\mu_e}{\mu_i}n_e \tilde{S}(u)\\
\nabla\cdot\left(n_e\nabla u-\tilde{T_e}\nabla n_e\right)=n_e \tilde S(u)
\end{eqnarray}
with boundary conditions, in compact form
\begin{eqnarray}
-\left[-\mbox{sgn}(q)n_q\nabla u-\nabla
  n_q\right]\cdot\normal=\nonumber\\
=\mbox{sgn}(q)\ a\ n_q\nabla u\cdot\normal+\delta_{qe}\left(\gamma_i \mbf{J_i}\cdot\normal+\frac{1}{4}n_q v_{\mbox{th}}\right)
\end{eqnarray}
where we have defined a non dimensional electron temperature
$\tilde{T_e}=T_e/T$, and we simply set $\tilde S(u)=S(V_0 u)$.
In the above equations, $q$ indicates the charge carrier, \ie, $i$ or $e$, and $a=1$ if
$-\mbox{sgn}(q)\nabla u\cdot\normal > 0$ and $a=0$ otherwise.
The last term starting with the Kronecker delta $\delta_{qe}$ applies
only to the electrons on ``cathode-like'' boundary conditions, and
is neglected elsewhere. 
\section{Numerical results}
Let us now show some numerical results obtained for some simple
configurations of the chamber in 1D and 2D. All the simulations were
performed with Comsol$^{\mbox{\textregistered}}$
Multiphysics~\cite{comsol2005comsol}, by using a modified
convection-diffusion model, for the two current densities, coupled with an
electrostatic model (Poisson equation) for the electric potential. 
\subsection{One dimensional model}
\begin{figure*}[htb]
\begin{center}
\begin{picture}(0,0)%
\includegraphics{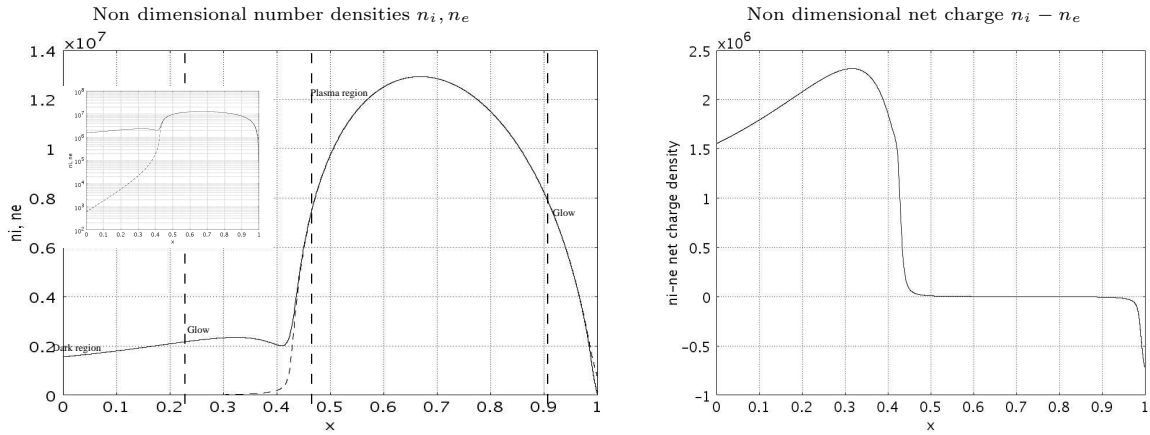}%
\end{picture}%
\setlength{\unitlength}{4144sp}%
\begingroup\makeatletter\ifx\SetFigFont\undefined%
\gdef\SetFigFont#1#2#3#4#5{%
  \reset@font\fontsize{#1}{#2pt}%
  \fontfamily{#3}\fontseries{#4}\fontshape{#5}%
  \selectfont}%
\fi\endgroup%
\begin{picture}(7350,2596)(271,-1951)
\put(1065,546){\makebox(0,0)[lb]{\smash{{\SetFigFont{7}{8.4}{\familydefault}{\mddefault}{\updefault}{\color[rgb]{0,0,0}Non dimensional number densities $n_i, n_e$}%
}}}}
\put(4980,546){\makebox(0,0)[lb]{\smash{{\SetFigFont{7}{8.4}{\familydefault}{\mddefault}{\updefault}{\color[rgb]{0,0,0}Non dimensional net charge $n_i-n_e$}%
}}}}
\end{picture}%
\caption{Left: non dimensional number density of ions and electrons
  (dashed). In the insert, same plot in logarithmic scale. As already
  shown in the plot of the electric field, the main expected features
  such as the ``cathode dark space'', the ``negative glow'', and the
  ``positive glow'' are evident. In the dark space, the number of ions
  is greatly larger than the number of electrons, but not yet
  maximum. A maximum is reached in the glow region. The plasma
  region is neutral, as the number of ions balances exactly the number
of electrons (overlapping curves). In the positive glow region, the
number of electrons exceeds the number of ions and the plasma in
negatively charged. Right: net charge density $\rho=n_i-n_e$. The
existence of the four different regions is evident as the 
net charge density changes drastically from one region to the other.}
\label{fig:1d_ni_ne}
\end{center}
\end{figure*}

In order to be able to interpret correctly the results obtained in 2D,
particularly in the configuration with two electrodes, let us start
with showing some results obtained in 1D. In this case, our domain is
simply a straight line extending between the cathode, on the left, and the
anode, on the right. The steel chamber in this case has no particular
role and, consequently, the boundary conditions greatly simplify as
there is no influence of direction of the electric field on the
surface. In spite of this drastic reduction of the model, as we will
see, the results already contain an important part of the information
given by the 2D case, in more realistic configurations. Let us then
start with a plot of the electric potential and electric field norm.
\begin{figure*}[htb]
\begin{center}
\begin{picture}(0,0)%
\includegraphics{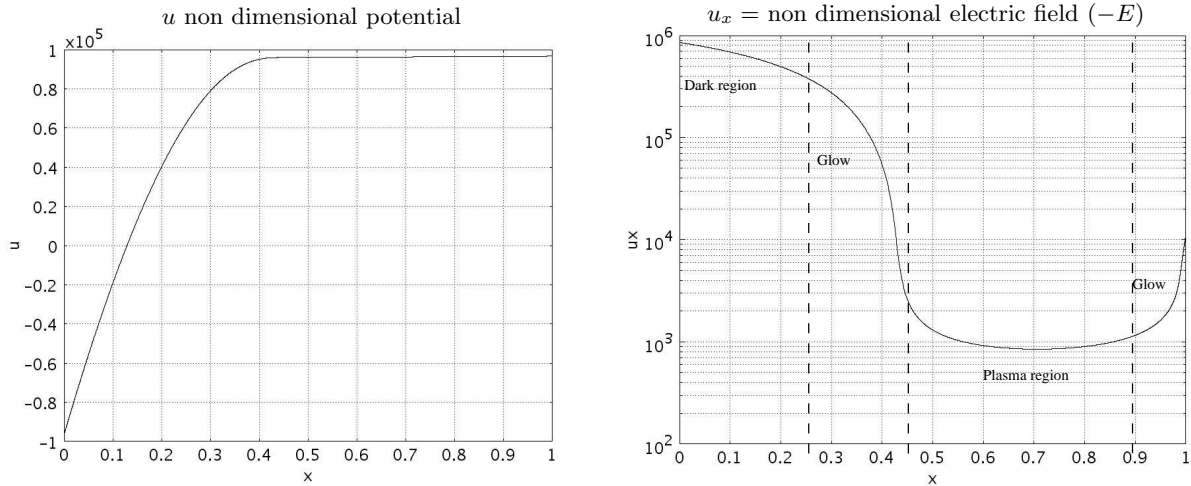}%
\end{picture}%
\setlength{\unitlength}{4144sp}%
\begingroup\makeatletter\ifx\SetFigFont\undefined%
\gdef\SetFigFont#1#2#3#4#5{%
  \reset@font\fontsize{#1}{#2pt}%
  \fontfamily{#3}\fontseries{#4}\fontshape{#5}%
  \selectfont}%
\fi\endgroup%
\begin{picture}(7522,2994)(402,-2773)
\put(4816, 74){\makebox(0,0)[lb]{\smash{{\SetFigFont{9}{10.8}{\familydefault}{\mddefault}{\updefault}{\color[rgb]{0,0,0}$u_x=$ non dimensional electric field ($-E$) }%
}}}}
\put(1553, 55){\makebox(0,0)[lb]{\smash{{\SetFigFont{9}{10.8}{\familydefault}{\mddefault}{\updefault}{\color[rgb]{0,0,0}$u$ non dimensional potential}%
}}}}
\end{picture}%
\caption{Non dimensional electric potential (left) and norm of the
  electric field (right). The dimensional
  potential difference between anode and cathode is set at 5000
  V. The norm of the electric field shows clearly the main features
  expected, such as the ``cathode dark space'', the negative and
  positive glow and a plasma region, with low electric field.}
\label{fig:1d_u_ux}
\end{center}
\end{figure*}
In Fig.~\ref{fig:1d_u_ux} we represent the non dimensional electric potential $u$
(left) and its spatial derivative $u_x$, \ie, the electric field (with
a positive sign), in logarithmic scale. The
electric potential clearly shows two main regimes of the solution. In
the first regime corresponding to a region
extending between the cathode and the center of the domain,
the electric potential grows steeply. Next to this, in a region extending from the
center to the other end of the domain, at the anode, the
electric potential has already reached its boundary  value. However, when we plot the electric field (right), in
logarithmic scale, we actually observe a very rich behaviour, showing
all the main expected features. First of all, starting from the
cathode (left), the ``cathode dark space'' extends for roughly one
fourth of the domain. The dark space is a region relatively poor of
ions (compared to the plasma region), where the charged particles are being
accelerated by a 
strong electric field. In the chamber, this small region appears
clearly as a dark space extending all along the cathode. Starting
from the right end of the dark space, up to almost a half of the
domain, we can observe the ``negative glow''. In this region, a large
number of highly energetic ions (accelerated in the dark space by the
electric field), release their energy in form of light. In the
experimental setting, one would see a vivid purple zone all around the
cathode. Next to the negative glow is a ``plasma region''. In this
region, almost nothing happens, in the sense that, in spite of the
large number of charged particles (see Fig.~\ref{fig:1d_ni_ne}), the
electric field here is so small that the energy density turns out to be very
low. 
Furthermore, the number of ions and electrons perfectly balance,
thus making the plasma region practically neutral (while the cathode
dark space and the negative glow are positively charged,
Fig.~\ref{fig:1d_ni_ne}-right). 
The other end of the domain, confined by the anode, is occupied by
the positive glow (and the anode dark space). In this region the electric field increases again,
although not as much as in the cathode dark space. However, as shown
in Fig.~\ref{fig:1d_ni_ne}, in this region, the electrons
exceed the ions, making the plasma negatively charged.

The representation of the net charge of the plasma, in
Fig.~\ref{fig:1d_ni_ne}-right confirms all our assertions. The cathode
dark space, is positively charged, and extends from the cathode up to
roughly 0.25. In the negative glow region, the net charge reaches its
maximum and then decreases steeply towards the plasma region, which is
neutral as a consequence of the perfect overlapping of the number
densities of ions and electrons. Finally, the positive glow shows a
negatively charged atmosphere, as the number of electrons exceeds the
number of ions in the nearby of the anode. Overall,
Fig.~\ref{fig:1d_ni_ne} shows that the plasma is positively charged
as the total number of ions exceeds the total number of electrons,
except in the nearby of the anode.
\subsection{Two dimensional model -- 1 electrode}
Let us now show some partial numerical results in the two dimensional
model. In order to correlate our results to the previous 1D case, let
us assume a simple configuration with one central electrode (cathode),
while the steel chamber acts like the anode and is kept at a neutral
potential. The cathode is set at a potential (dimensional $V$) of
-3000 V with respect to the steel chamber. 

\begin{figure*}[htb]
\begin{center}
\epsfig{file=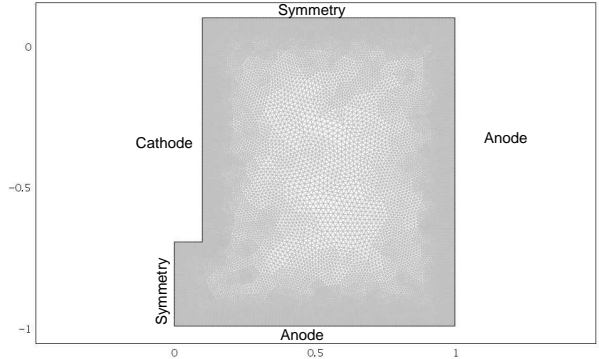,width=8cm}
\caption{Representation of the two dimensional mesh used in the
  cylindrically symmetric case with one electrode. The domain is
  delimited on the left by the cathode, while the steel chamber
  (bottom and right) constitutes the anode. At the top we impose a
  symmetry condition as on the central axis.}
\label{fig:mesh_1elettrodo}
\end{center}
\end{figure*}

In this simple case (Fig.~\ref{fig:mesh_1elettrodo}), we can use the cylindrical symmetry of the
domain, and reduce the real 3D case to a simpler 2D model. 
In fact, one can reduce the integration domain to
just a half (we chose the lower half) of the vertical section of the
chamber, as the upper half is perfectly symmetric to the lower
part. 
%%%%%%%%%%%%%%%%%%%%%%%REFEREE 2 - P. 6%%%%%%%%%%%%%%%%%%%%%%%%%%%%%%%%%%
The following results were obtained with a FEM triangular mesh (Fig.~\ref{fig:mesh_1elettrodo})
consisting of 37456 non uniform elements. The system of equations was solved through an affine
invariant form of the damped Newton
method~\cite{comsol2005comsol,Deuflhard74}, implemented through a
direct linear solver UMFPACK~\cite{comsol2005comsol} included in Comsol$^{\mbox{\textregistered}}$
Multiphysics.
%%%%%%%%%%%%%%%%%END OF REFEREE PARAGRAPH%%%%%%%%%%%%%%%%%%%%%%%%%%%%%%  

\begin{figure*}[htb]
\begin{center}
\begin{picture}(0,0)%
\includegraphics{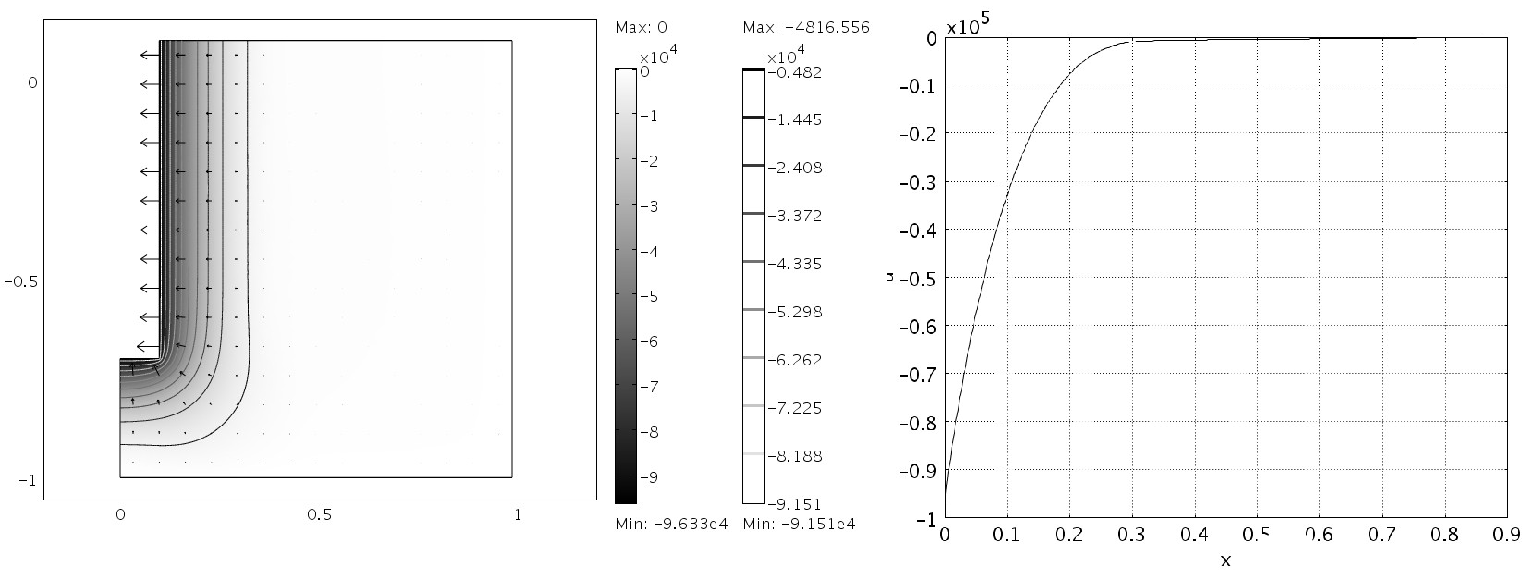}%
\end{picture}%
\setlength{\unitlength}{4144sp}%
\begingroup\makeatletter\ifx\SetFigFont\undefined%
\gdef\SetFigFont#1#2#3#4#5{%
  \reset@font\fontsize{#1}{#2pt}%
  \fontfamily{#3}\fontseries{#4}\fontshape{#5}%
  \selectfont}%
\fi\endgroup%
\begin{picture}(7155,3140)(631,-2661)
\put(6082,-2610){\makebox(0,0)[lb]{\smash{{\SetFigFont{8}{9.6}{\familydefault}{\mddefault}{\updefault}{\color[rgb]{0,0,0}Arc-length}%
}}}}
\put(1081,281){\makebox(0,0)[lb]{\smash{{\SetFigFont{8}{9.6}{\familydefault}{\mddefault}{\updefault}{\color[rgb]{0,0,0}Non  dimensional electric potential $u$}%
}}}}
\put(4750,297){\makebox(0,0)[lb]{\smash{{\SetFigFont{8}{9.6}{\familydefault}{\mddefault}{\updefault}{\color[rgb]{0,0,0}Non dim. electric potential $u$ (cross section)}%
}}}}
\end{picture}%
\caption{Left: Representation of the non dimensional electric potential $u$
  in the two-dimensional case with one central electrode. The
  potential increases steeply from the cathode into the domain,
  reaching quickly a value already close to the one imposed at the
  anode $u=0$. The arrows pointing inwards the cathode show the
  strength of the electric field in that area, compared to the plasma
  region, where they are invisible as the electric field is low.
 Right: Cross section obtained by drawing a horizontal line at the top
of the domain, \ie, on the horizontal symmetry axis. The plot shows
the perfect correspondence of this case, to the 1D case.} 
\label{fig:1elettrodo_u}
\end{center}
\end{figure*}

\begin{figure*}[htb]
\begin{center}
\begin{picture}(0,0)%
\includegraphics{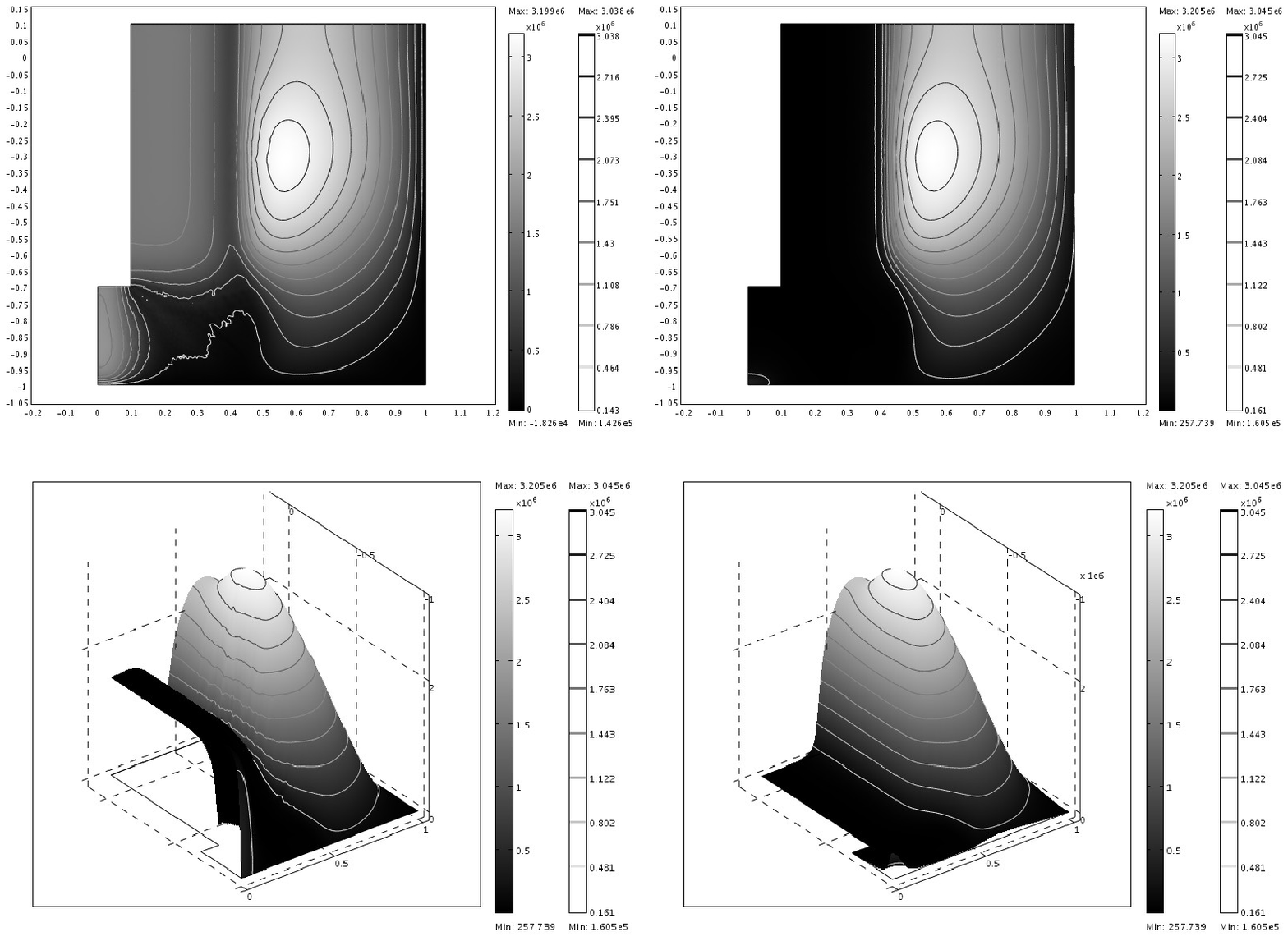}%
\end{picture}%
\setlength{\unitlength}{4144sp}%
\begingroup\makeatletter\ifx\SetFigFont\undefined%
\gdef\SetFigFont#1#2#3#4#5{%
  \reset@font\fontsize{#1}{#2pt}%
  \fontfamily{#3}\fontseries{#4}\fontshape{#5}%
  \selectfont}%
\fi\endgroup%
\begin{picture}(7508,5705)(451,-4908)
\put(692,674){\makebox(0,0)[lb]{\smash{{\SetFigFont{8}{9.6}{\familydefault}{\mddefault}{\updefault}{\color[rgb]{0,0,0}Non dimensional ions number density $n_i$}%
}}}}
\put(4600,674){\makebox(0,0)[lb]{\smash{{\SetFigFont{8}{9.6}{\familydefault}{\mddefault}{\updefault}{\color[rgb]{0,0,0}Non dimensional electrons number density $n_e$}%
}}}}
\end{picture}%
\caption{Left: Contour plot (top) and 3D surface (bottom) of the non dimensional number density of
ions (top). Right: Contour plot (top) and 3D surface (bottom) of the non
dimensional number density of electrons. As in the 1D model,
the two distributions overlap perfectly in the central area
(plasma region) of
the domain. Near the cathode (dark space and glow) the number of ions largely exceeds the
number of electrons, thus making the plasma overall, positively charged.}
\label{fig:1elettrodo_ni_ne}
\end{center}
\end{figure*}

As in the previous case, a plot of the non dimensional electric
potential (Fig.~\ref{fig:1elettrodo_u}), shows the two main regimes with the potential increasing
steeply from the cathode inside the region, and a wide plasma region,
where the electric potential has already reached a value which is very
close to the one imposed at the anode ($u=0$ in this case).
The similarity between the behaviour of the 1D case and the 2D
case with one central electrode is evident when we plot
the electric potential evaluated along a horizontal line drawn over
the symmetry axis that closes the domain at the top
(Fig.~\ref{fig:1elettrodo_u}-right). As in the previous
1D model, the two regimes (actually hiding the four
regions already shown in the 1D model) with the potential
increasing steeply proceeding from the cathode inwards the integration
domain, and the ``plasma region'' with an almost constant potential,
are clearly evident. A similar parallelism between the 1D model
and the cylindrically symmetric 2D one is evident when
we plot the number densities of ions $n_i$ and electrons $n_e$
(Fig.~\ref{fig:1elettrodo_ni_ne}). 
%The same conclusions as in the one dimensional case can be drawn from
%the observation of the distribution of the number densities of ions
%($n_i$) and electrons ($n_e$), in
%Fig.~\ref{fig:1elettrodo_ni_ne}. 
Again, as in the previous case, the
two distributions overlap perfectly in the ``plasma region'', while
near the cathode the number of ions is largely greater than the number
of electrons. Viceversa, near the anode (the steel chamber in this
case), the electrons exceed the ions, thus making the plasma locally
negative, but positive overall. Fig.~\ref{fig:1elettrodo_ni_ne} shows
also a clear resemblance with the number density of positive ions and
electrons reported by Boeuf in~\cite{boeuf1988tdm}. The main
differences between our case and theirs, besides, possibly, the value
of the scaling factors, is due probably due to the different choice of
boundary conditions for the potential of the steel chamber (linear in
their case, uniform in ours). 
\subsection{Two dimensional model -- two electrodes}
\begin{figure*}[htb]
\begin{center}
\epsfig{file=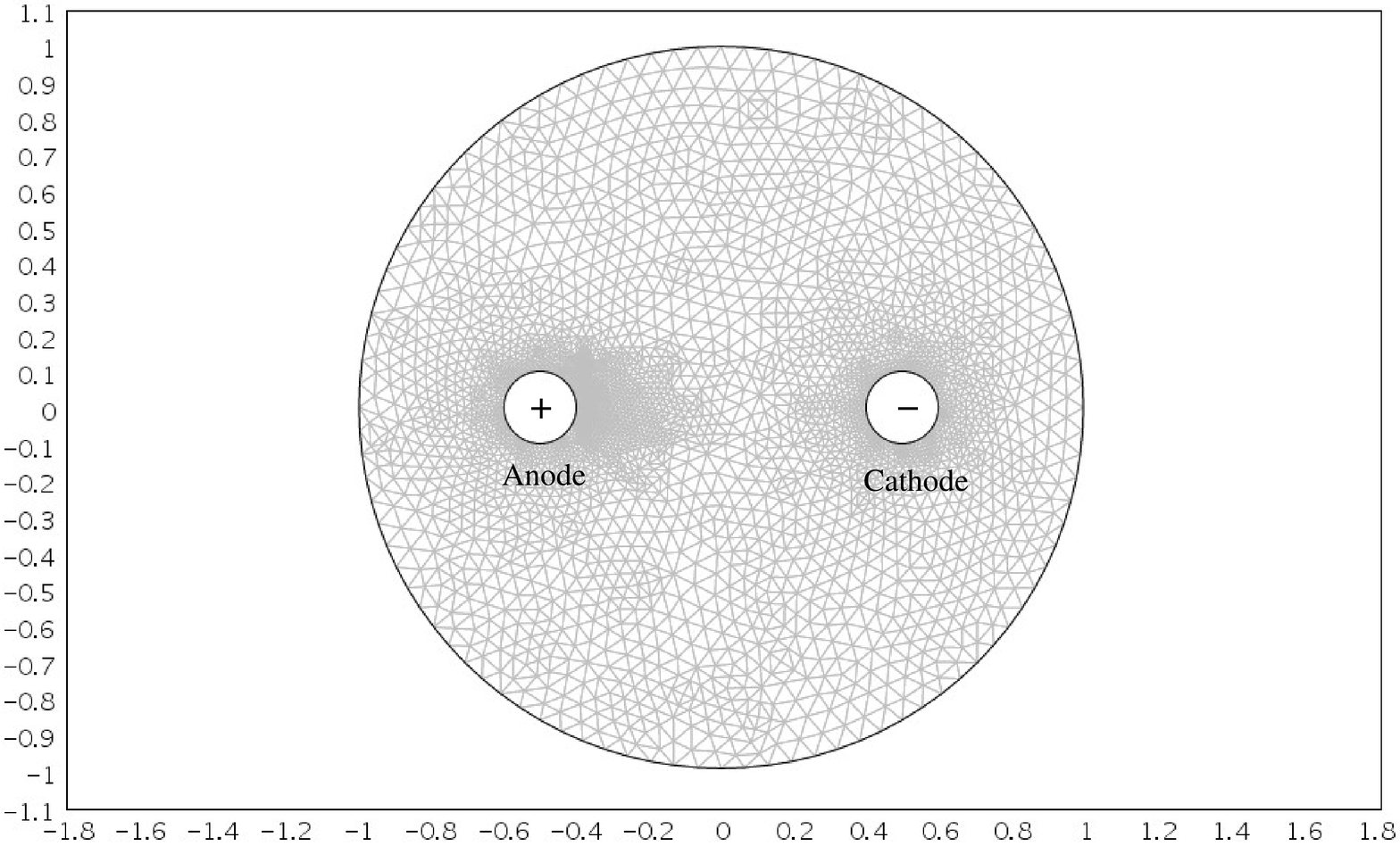,width=10cm}
\caption{Representation of the mesh used to solve the model in the two
dimensional case, obtained by cutting horizontally the original 3D
domain. The two cylindrical electrodes are represented by two circles,
the anode on the left, and the cathode on the right.}
\label{fig:2d_2elettrodi_mesh}
\end{center}
\end{figure*}

\begin{figure*}[htb]
\begin{center}
\begin{picture}(0,0)%
\includegraphics{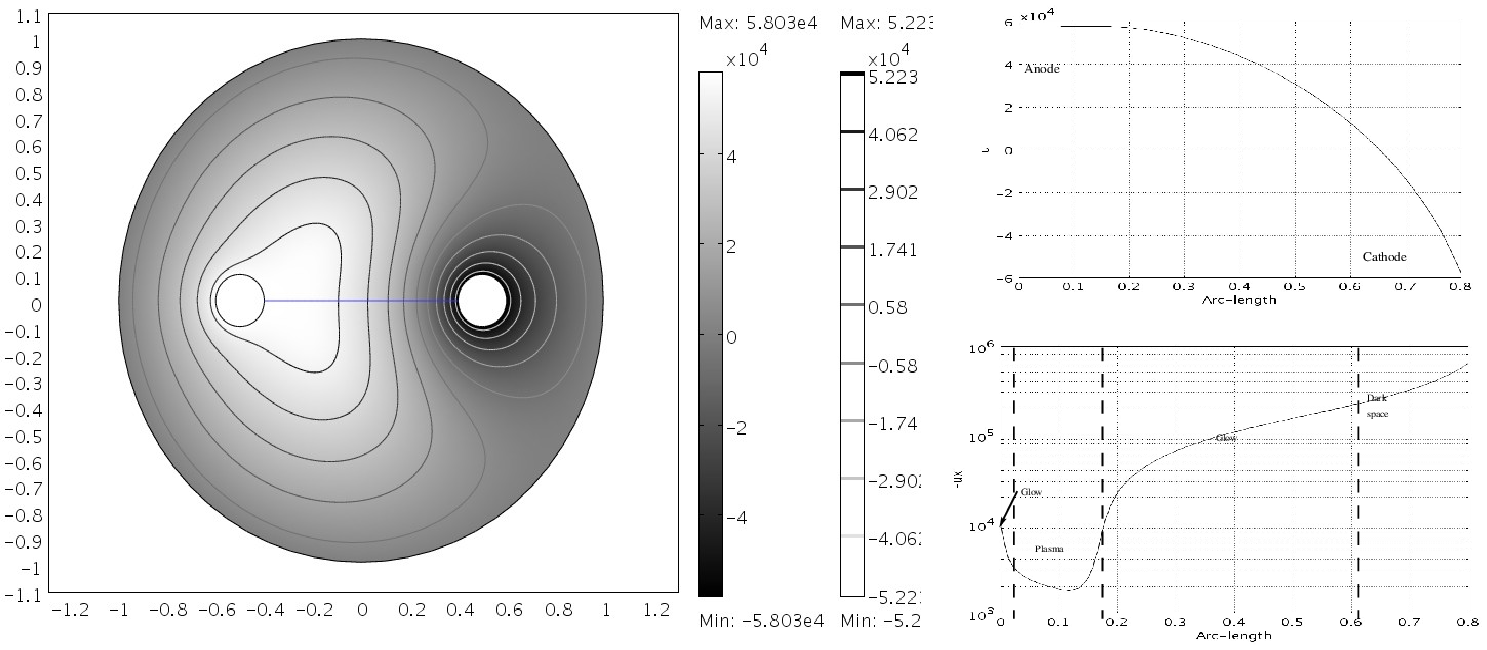}%
\end{picture}%
\setlength{\unitlength}{4144sp}%
\begingroup\makeatletter\ifx\SetFigFont\undefined%
\gdef\SetFigFont#1#2#3#4#5{%
  \reset@font\fontsize{#1}{#2pt}%
  \fontfamily{#3}\fontseries{#4}\fontshape{#5}%
  \selectfont}%
\fi\endgroup%
\begin{picture}(6885,3645)(451,-5911)
\put(5485,-4134){\makebox(0,0)[lb]{\smash{{\SetFigFont{5}{6.0}{\familydefault}{\mddefault}{\updefault}{\color[rgb]{0,0,0}Electric field norm, cross section}%
}}}}
\put(5491,-2536){\makebox(0,0)[lb]{\smash{{\SetFigFont{5}{6.0}{\familydefault}{\mddefault}{\updefault}{\color[rgb]{0,0,0}Electric potential $u$, cross section}%
}}}}
\put(1306,-2536){\makebox(0,0)[lb]{\smash{{\SetFigFont{6}{7.2}{\familydefault}{\mddefault}{\updefault}{\color[rgb]{0,0,0}Non dimensional electric potential $u$}%
}}}}
\end{picture}%
\caption{Non dimensional potential u in contour plot (left) and along
  the cross section between the electrodes sketched as the horizontal
  segment between the electrodes (right/top). Both representations of
  the potential show the dominance of the area at positive potential,
  bounded by the anode,
  compared to the area at negative potential bounded by the
  cathode. Right/bottom, non dimensional electric field (norm) along
  the same cross section as above. The electric field shows again the
  typical features as the previous cases, such as the dark space, the
  positive and negative glows, and the plasma region.}
\label{fig:2d_2elettrodi_potential}
\end{center}
\end{figure*}

Let us now approach the case with the widest industrial interest, as
it resembles, although yet in a two dimensional representation, the
real plant, in the configuration with two electrodes and
an alternate electric field. 
As the cylindrical symmetry is lost, due to the introduction of the
second electrode, none of the possible two dimensional reductions can represent the
real three dimensional model. However, a horizontal slice of the three
dimensional domain (Fig.~\ref{fig:2d_2elettrodi_mesh}), as we will see, already gives some useful insights of what happens in
the central (in height) area of the electrodes of the full three
dimensional case, where we can assume some sort of limited translational
invariance. The FEM mesh consisted of 6560 non uniform standard
triangular elements. 
As in the previous cases in one or two dimensions, the 2D model
with two electrodes shows the typical expected
behaviour. Once again, the plot of the electric potential (contour
plot in Fig.~\ref{fig:2d_2elettrodi_potential}-left) shows a dominance
of the region extending from the anode inwards, where the potential is
roughly constant. A plot of the potential evaluated along the
orthogonal cross section between the two electrodes (sketched in
Fig.~\ref{fig:2d_2elettrodi_potential}-left as a horizontal segment
between the electrodes) shows that the two-dimensional case resembles
accurately the previous 1D and 2D models. 
As above, the electric potential remains constant in the nearby of the
anode and decreases steeply only as we get closer to the cathode. The
plot of the norm of the electric field (non dimensional), in
Fig.~\ref{fig:2d_2elettrodi_potential}-right/bottom, shows again the
features already observed in the 1D model, \ie, the dark
space, the two glows and the plasma region, although the extension of
the regions is now completely different from the previous cases. 

\begin{figure*}[htb]
\begin{center}
\begin{picture}(0,0)%
\includegraphics{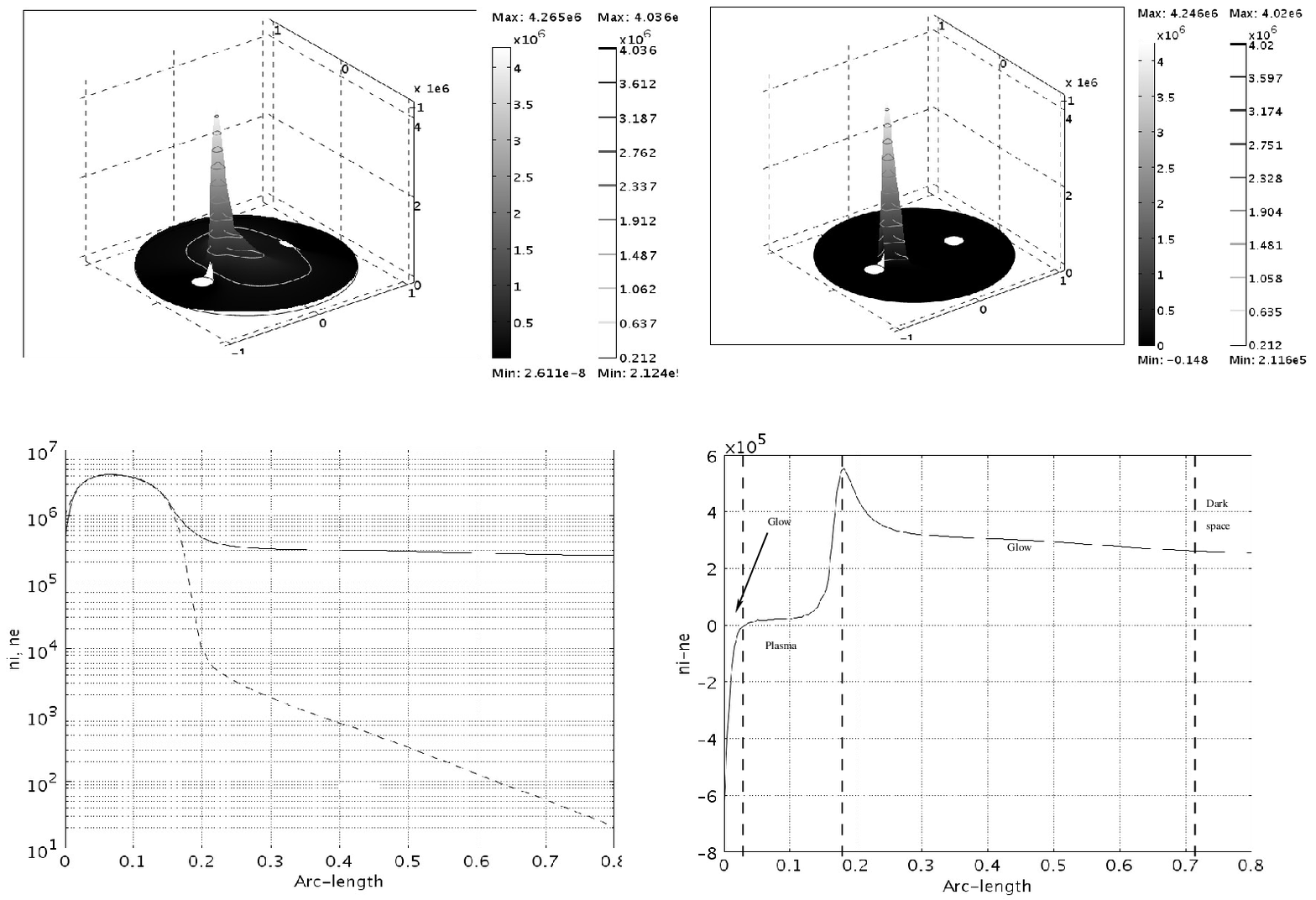}%
\end{picture}%
\setlength{\unitlength}{4144sp}%
\begingroup\makeatletter\ifx\SetFigFont\undefined%
\gdef\SetFigFont#1#2#3#4#5{%
  \reset@font\fontsize{#1}{#2pt}%
  \fontfamily{#3}\fontseries{#4}\fontshape{#5}%
  \selectfont}%
\fi\endgroup%
\begin{picture}(7470,5318)(541,-5063)
\put(4693,134){\makebox(0,0)[lb]{\smash{{\SetFigFont{7}{8.4}{\familydefault}{\mddefault}{\updefault}{\color[rgb]{0,0,0}Non dimensional electrons number density $n_e$}%
}}}}
\put(4693,-2224){\makebox(0,0)[lb]{\smash{{\SetFigFont{7}{8.4}{\familydefault}{\mddefault}{\updefault}{\color[rgb]{0,0,0}Non dimensional net charge density $n_i-n_e$}%
}}}}
\put(1151,-2224){\makebox(0,0)[lb]{\smash{{\SetFigFont{7}{8.4}{\familydefault}{\mddefault}{\updefault}{\color[rgb]{0,0,0}Non dimensional number densities $n_i$ and $n_e$}%
}}}}
\put(1483,144){\makebox(0,0)[lb]{\smash{{\SetFigFont{7}{8.4}{\familydefault}{\mddefault}{\updefault}{\color[rgb]{0,0,0}Non dimensional ions number density $n_i$}%
}}}}
\end{picture}%
\caption{Non dimensional density of ions (top/left) and electrons
  (top/right) in three dimensional representation. The two
  distributions are strongly peaked in the plasma region. However, the
  two peaks overlap almost exactly, as clear from the cross section
  plot (bottom/left) and from the cross section plot of the net charge
  density $n_i-n_e$ (bottom/right). The net charge density
  (bottom/right) shows also the typical behaviour encountered in the
  previous cases. The plasma region is roughly neutral, while the
  cathode dark space and the negative glow are positively charged. The
positive glow is negatively charged, as the number of electrons
exceeds the number of ions.}
\label{fig:2d_2elettrodi_ni_ne}
\end{center}
\end{figure*}

In spite of the orthogonality between the two projections (horizontal
in this case vs vertical in the cylindrically symmetric one), the great
resemblance of this 2D case with the previous cylindrically 
symmetric model, is
clearly evident also when we plot the number density of ions $n_i$ and
$n_e$ (Fig.~\ref{fig:2d_2elettrodi_ni_ne}). However, in this case, the
number density shows some extreme features that were not so evident
in the previous cases.
Fig.~\ref{fig:2d_2elettrodi_ni_ne}-top shows that both the number
density of ions (left) and electrons (right) are strongly peaked in
the plasma region. In fact, as before, the two peaks are exactly
overlapping, as it is evident from the cross section plot of the two
densities (bottom-left) and of the net charge $n_i-n_e$
(bottom-right), where we see how, in the plasma region, the net charge
is, once again, almost exactly null. As before, the plasma is then positively
charged in the cathode dark space and the cathode glow, and
negatively charged in the anode glow.

\begin{figure*}[htb]
\begin{center}
\begin{picture}(0,0)%
\includegraphics{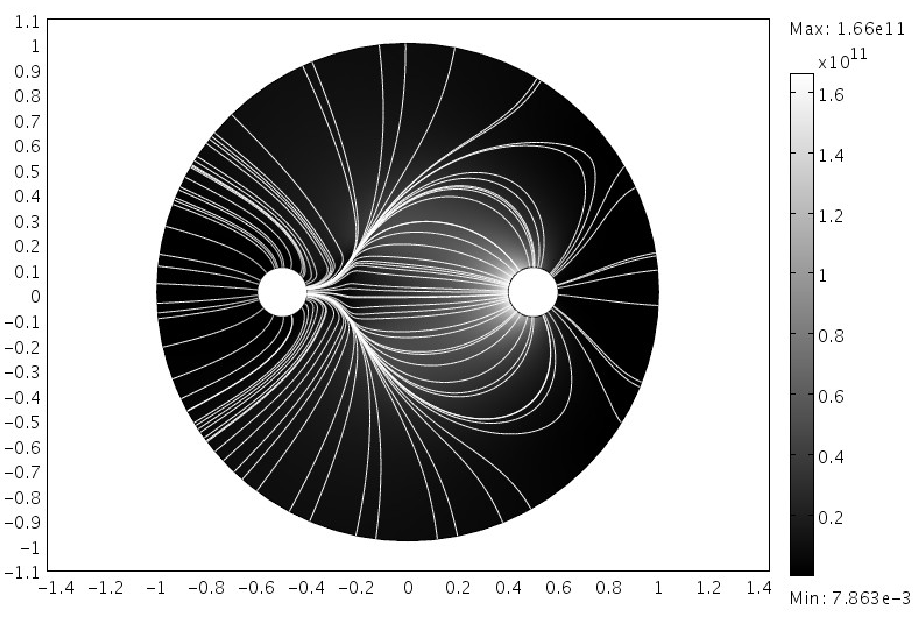}%
\end{picture}%
\setlength{\unitlength}{4144sp}%
\begingroup\makeatletter\ifx\SetFigFont\undefined%
\gdef\SetFigFont#1#2#3#4#5{%
  \reset@font\fontsize{#1}{#2pt}%
  \fontfamily{#3}\fontseries{#4}\fontshape{#5}%
  \selectfont}%
\fi\endgroup%
\begin{picture}(4260,3195)(1486,-5011)
\put(4633,-2978){\rotatebox{270.0}{\makebox(0,0)[lb]{\smash{{\SetFigFont{6}{7.2}{\familydefault}{\mddefault}{\updefault}{\color[rgb]{0,0,0}Anode-like}%
}}}}}
\put(2841,-2010){\makebox(0,0)[lb]{\smash{{\SetFigFont{6}{7.2}{\familydefault}{\mddefault}{\updefault}{\color[rgb]{0,0,0}Ions current density$J_i$ }%
}}}}
\put(2116,-3616){\rotatebox{90.0}{\makebox(0,0)[lb]{\smash{{\SetFigFont{6}{7.2}{\familydefault}{\mddefault}{\updefault}{\color[rgb]{0,0,0}Cathode-like}%
}}}}}
\end{picture}%
\caption{Ions density flux $J_i$ and streamlines of the electric
  field. The current flows, as expected, mainly between the two
  electrodes, with a high density of ions flux in the vicinity of the
  cathode. However, some electric field streamlines close up on the
  steel chamber. The steel chamber therefore contributes to the total
  current flow within the chamber, due to a partial coupling
  between the electrodes and the chamber itself. In the area where
  the streamlines connect the chamber to the anode, the chamber
  behaves like a cathode, while in the area where the streamlines
  connect to the cathode, the chamber behaves like the anode.}
\label{fig:2d_2elettrodi_current}
\end{center}
\end{figure*}

Finally, let us go back for a moment to the boundary conditions
imposed in to the steel chamber in
Eqs.~\eqq{eq:ion_chamber}{eq:electron_chamber}. There, we said, we
have to impose a different condition, depending on the sign of the
electric field on the surface. In other words, we assumed that,
somehow, the steel chamber contributes to the total current crossing
the chamber, behaving, in some parts, anode-like, and in others,
cathode-like. An insight of this effect, which is clearly rather
weak, but yet present, can be obtained by showing a plot of the stream
lines of the current density of ions
$J_i$. Fig.~\ref{fig:2d_2elettrodi_current} shows, as expected, that
the ions flux (and similarly the electrons flux, not shown) is
concentrated within the region between the electrodes. However, some
(non negligible)
density is flowing also in the area around the two electrodes, between
the electrodes and the steel chamber. There, the electric field
streamlines close between the electrodes and the steel chamber.
In other words, the steel chamber, through a partial coupling with
the electrodes, contributes to the flow of electric current. In
particular, the steel chamber behaves like an anode in the area where
the streamlines connect it to the cathode, while it behaves like a
cathode, where the streamlines connect it to the anode (most of the
surface in this case). As mentioned, the contribution of this partial
coupling to the total amount of current is fairly low. However, this
contribution exists and confirms that the boundary conditions must
include it somehow. Furthermore, this suggests that also a coupling of
different nature, for instance, due to residual capacity effects, is
possible between the steel 
chamber and the electrodes. This type of coupling was completely
neglected in our case, since we assumed a static field, however, this
result shows that, in case of alternate current, it could contribute
even further to the passage of electric current through the plant.
\section{Conclusions}
We analyzed the behaviour of plasma atmosphere in a PVD industrial
plant by means of a mathematical model involving the electric
potential and the number density of ions and electrons. The drastic
simplifications we made in setting the mathematical model, allowed us
to solve it numerically in a fairly straightforward way. The main
difficulty of the model, as already mentioned in the literature,
concerns the setting of the boundary conditions. In fact, previous
results show that the usual Neumann or Dirichlet boundary conditions
are too simple to give physically meaningful results. For this reason,
we made an intermediate choice of mixed boundary conditions, that take
into account of, at least, secondary and thermal emission of
electrons. However, as in our case the plant configuration consists of
two electrodes and the steel chamber, the boundary conditions turn out to depend on the solution
inside the domain. The three equations of the model are therefore
coupled 
in a strongly non linear way. In spite of this, for fairly simple
geometries, such as the shown two dimensional cut, and the corresponding
full three dimensional case (not included), the numerical solution of the model is still possible. In fact,
the two dimensional model with two electrodes show a great
resemblance with the one dimensional case and the cylindrically
symmetric two dimensional case. The three
analyzed cases show a similar behaviour of the electric potential,
with evidence of two different regimes. A plasma region is clearly
evident with almost constant potential, bounded by the
anode. A highly energetic glow region with the potential increasing
steeply, extends between the plasma and the cathode. 

Similar considerations can be made, concerning the number densities of
ions and electrons. In all the shown cases (and similarly it happens
in the three dimensional case, although that is rather more difficult
to visualize) the number density of ions and electrons overlap almost
exactly in the plasma region. In the cathode dark space and positive
glow, the ions greatly dominate, while in the negative glow (close to
the anode), the number of electrons exceeds the number of ions. This
behaviour was again expected, since it was both observed
experimentally and simulated by a mathematical model for the one
dimensional case (such as the glass tube~\cite{plasmas}) and some simple
configurations of the 2D model~\cite{boeuf1988tdm}. As far as we know,
this is the first case where a full two-dimensional model, with no
rotation invariance, is solved numerically and fully reported. Furthermore, as far as we
know, it is the first time that a clear comparison between the 1D
model and more complex 2D models, either with one or two electrodes,
is made. Clearly, this is not
conclusive, as the full three dimensional was not analyzed in
detail. However, it is the first time that a published numerical model
actually takes into the account of the real industrial configuration
of an alternate current (AC) plant.

In terms of the application of our results in an
industrial contest, one should try to correlate these theoretical
considerations with some experimental results of the deposition
process, in order to show the influence of the different conditions of
the plasma on the polymeric film formed. 
%In this way one could see whether
%the best results are obtained in the plasma region or closer to the
%electrodes, or, as in the actual configuration, outside the region
%between the two electrodes and near the steel chamber. 
Clearly, only partial
considerations can be made by a discussion of our
numerical results. One expects, for instance, the plasma region to be fairly quiet,
in terms of electric energy, as in spite of the great number of
charged particles (both ions and electrons reach their absolute maximum in the
plasma region), the electric field is almost null. 
%The charged
%particles are thus not accelerated by the field in the plasma
%region. 
Conversely, in the negative glow region, a large number of
ions (compared to the electrons) are strongly accelerated by a strong
electric field. This is the most energetic region of the whole domain,
and one expects a rather dramatic behaviour of the plasma. For
example, in the glow region (or the cathode dark space), one expects a
strong ``sputtering effect'' due to the ions hitting the surface of
substrate to be treated and, possibly, removing the aluminium film. That
is clearly a phenomenon that one would want to avoid as much as
possible, when choosing where to pose the pieces to be treated within the
chamber. However, at this stage, \ie, without a systematic
experimental campaign it is hard to give any further interpretation of
the model, except for very general considerations, concerning the real
industrial process of polymerization and coatings. In spite of this,
we believe our model can be a good starting point for a systematic
investigation of the real process. In order to interpret
correctly our results and guide the industrial process, one should,
possibly, not only compare them with experiments and direct
observations, but also with the results of a simplified polymerization
model. One could then correlate the results from the two models
with observations, and draw some conclusions about the real process of
coating. For instance, a deeper investigation of the polymerization
process could suggest that a highly energetic region is favorable in
terms of speed and effectiveness of the polymerization. 

Although this project of investigation is just at the
beginning, we believe that our model marked a good step forward in the
direction of a more systematic and scientific approach to the
optimization of a commonly used, but not yet fully understood,
industrial process.

%\begin{acknowledgements}
{\bf Acknowledgements}
 The authors are deeply grateful to Mr. Alessandro
  Carletti, Dr. Massimo Lasagni and Dr. Fabiano Remediotti from Galileo
  Vacuum Systems for the given support and the helpful discussions
  throughout the research, and to Galileo Vacuum Systems s.r.l. for funding the
  project and providing most of the technical literature.
%\end{acknowledgements}

\bibliographystyle{plain}
\bibliography{vpd}

\end{document}